\documentclass[aps,amsmath,amssymb,prb,twocolumn,showpacs]{revtex4}

\usepackage{bm}
\usepackage{epsfig}

\newcommand{\2}{{(2)}}
\newcommand{\3}{{(3)}}

\newcommand{\coma}{preprint cond-mat/}

\newcommand{\w}{\omega}

\newcommand{\rmi}{{\textrm i}}
\newcommand{\rmf}{{\textrm f}}

\newcommand{\delq}{\delta_{\{q\}}}
\newcommand{\B}{B^\3_{\{q\}}}
\newcommand{\E}{E^\3_{\{q\}}}
\newcommand{\F}{F^\3_{\{q\}}}

\newcommand{\trel}{t_\textrm{rel}}
\newcommand{\lth}{\ell_\textrm{th}}
\newcommand{\lqu}{\ell_\textrm{qu}}
\newcommand{\Eth}{E_\textrm{th}}
\newcommand{\Equ}{E_\textrm{qu}}

\newcommand{\cO}{{\mathcal{O}}}
\newcommand{\cZ}{{\mathcal{Z}}}
\newcommand{\cU}{{\mathcal{U}}}
\newcommand{\cF}{{\mathcal{F}}}

\newcommand{\hB}{\hat B^\3_{\{q\}}}
\newcommand{\hT}{\hat T}
\newcommand{\hU}{\hat U^\3_{\{q\}}}
\newcommand{\hrho}{{\hat \rho}}
\newcommand{\hhbar}{\hat \hbar}
\newcommand{\hht}{{\hat t}}
\newcommand{\htp}{{\hat t_+}}
\newcommand{\htm}{{\hat t_-}}
\newcommand{\hw}{{\hat \omega}}
\newcommand{\hnu}{{\hat \nu}}
\newcommand{\hmu}{{\hat \mu}}
\newcommand{\hmucl}{{\hat \mu}_{2,\textrm{cl}}^\3}
\newcommand{\hmuad}{{\hat \mu}_{2,\textrm{ad}}^\3}
\newcommand{\hmuhT}{{\hat \mu}_{2,\textrm{hT}}^\3}
\newcommand{\hmuhf}{{\hat \mu}_{2,\textrm{hf}}^\3}

\newcommand{\tX}{{\widetilde X}}

\newcommand{\tP}{{\widetilde P}}

\newcommand{\s}[1]{{s_{#1}}}
\newcommand{\ta}[1]{{\tau_{#1}}}
\newcommand{\q}[1]{q_{#1}}
\newcommand{\hq}[1]{\hat q_{#1}}
\newcommand{\U}[1]{U_{q_{#1}}}
\newcommand{\nuq}[1]{\nu_{q_{#1}}}

\begin{document}

\title{Quantum Brownian motion in ratchet potentials}

\author{Stefan Scheidl}

\affiliation{Institut f\"ur Theoretische Physik, Universit\"at zu
K\"oln, Z\"ulpicher Stra\ss e 77, D-50937 K\"oln, Germany}

\author{Valerii M. Vinokur}

\affiliation{Materials Science Division, Argonne National Laboratory,
  9700 S. Cass Avenue, Argonne, Il 60439}

\date{\today}

\begin{abstract}
  We investigate the dynamics of quantum particles in a ratchet
  potential subject to an ac force field.  We develop a perturbative
  approach for weak ratchet potentials and force fields.  Within this
  approach, we obtain an analytic description of dc current
  rectification and current reversals.  Transport characteristics for
  various limiting cases -- such as the classical limit, limit of high
  or low frequencies, and/or high temperatures -- are derived
  explicitly.  To gain insight into the intricate dependence of the
  rectified current on the relevant parameters, we identify
  characteristic scales and obtain the response of the ratchet system
  in terms of scaling functions.  We pay a special attention to
  inertial effects and show that they are often relevant, for example,
  at high temperatures.  We find that the high temperature decay of
  the rectified current follows an algebraic law with a non-trivial
  exponent, $j\propto T^{-17/6}$.
\end{abstract}


\maketitle

\section{Introduction}

Ratchets have attracted a considerable recent interest because of
their paradigmatic role as microscopic transport devices (for review
articles, see e.g. Refs.  \onlinecite{Astumian97,Bier97,Reimann00}).
Applications range from microscale electronics -- including the
photogalvanic effect,\cite{Belinicher+80} transport in quantum
dots\cite{Linke+98,Linke+98:err} and antidot arrays\cite{Lorke+98} --
over Josephson junctions \cite{Zapata+96,Falo+99,Weiss+00,Goldobin+01}
and vortex matter\cite{Olson+01} to cell biology.\cite{Julicher+97} At
the same time, ratchets are of fundamental theoretical interest since
they represent one of the simplest nonequilibrium systems.

The analysis of ratchet systems reaches back quite some time before
Feynman drew the attention of a wide audience to such systems in his
lectures where he discussed the possibility to employ ratchets as heat
engines.\cite{Feynman63} Subsequently, researches in ratchets have
been progressing steadily, in parallel in different scientific
communities, until the explosive outburst in theoretical and
experimental interest in the 1990s \cite{note:hist} had occurred.

In this paper, we report on the analytic progress in the study of the
so called tilting ratchets, where the combination of an asymmetric
static potential with an unbiased ac force and a coupling to a heat
bath leads to current rectification.  Past theoretical studies of this
ratchet type focused on the classical massless case
\cite{Magnasco93,Bartussek+94} and have revealed the \textit{current
  reversal} phenomenon, i.e., the possibility that the direction of
the rectified current reverses its direction when model parameters
such as the frequency or amplitude of the ac current are changed.  The
inclusion of a finite mass of the particles has shown that it may give
rise even to \textit{multiple} current reversals.\cite{Jung+96}
Further extensions accounted for the quantum nature of particles and
of the bath. Quantum fluctuations were found to provide an additional
source of current reversals.\cite{Reimann+97,Yukawa+97,Tatara+98}

In essence, the direction and amplitude of the current turned out to
be very sensitive to the various system parameters.  While this
dependence makes ratchets valuable for applications -- such as devices
that can separate particles of different species -- it still lacks a
satisfying theoretical understanding.  Analytic approaches can give
insight into this problem. However, even the single-particle problem
is already so complex that analytic approaches can be advanced only in
limiting cases, such as the adiabatic limit\cite{Reimann+97} or the
deterministic limit.\cite{Jung+96,Mateos00}

In our paper we develop a perturbative approach valid for weak ratchet
potentials and weak driving forces, which covers a wide range of
practical applications.  Within this perturbative approach we are able
to capture all prominent phenomena including multiple current
reversals. This approach provides a unified framework for deriving and
understanding the dependence of the rectified current on the particle
mass, temperature, friction coefficient, and frequency of the driving
force.  We pay particular attention to the role of inertial effects
and show that they lead to a substantial current enhancement even in
the high-temperature limit.

In Sec. \ref{sec.model} we specify the model and establish a
path-integral formulation as analytic framework.  The perturbative
scheme is developed in Sec. \ref{sec.pert}.  In Sec. \ref{sec.mu1} we
briefly demonstrate that the linear mobility can be conveniently
obtained from this approach and that results for special cases known
in the literature are reproduced.  However, ratchet effects can be
obtained only in nonlinear response.  The leading nonlinear mobility
is calculated and evaluated for various limiting cases in Sec.
\ref{sec.mu2}.  We conclude with a discussion of our approach and
results in Sec. \ref{sec.disc}.  Technical details of our calculations
are presented in appendices.

\section{Model}
\label{sec.model}

We consider a quantum particle of the mass $m$ in a stationary ratchet
potential $U(x)$.  In addition, we impose an ac driving force $F(t)$
which is chosen to be unbiased, i.e., to vanish upon time averaging.
Following Caldeira and Leggett,\cite{Caldeira+83} we couple the
particle linearly to a bath of harmonic oscillators at temperature
$T$.  This bath simultaneously provides friction and a fluctuating
force for the particle.  For simplicity, we assume a linear spectral
distribution of these oscillators, giving rise to Ohmic dissipation.
In the classical limit, the particle coordinate $x(t)$ follows the
equation of motion
\begin{eqnarray}
  m \ddot x(t) =  - U'(x(t)) +  F(t) - \eta \dot x(t) + \xi(t)
\end{eqnarray}
with a friction coefficient $\eta$ and a Gaussian thermal noise
$\xi(t)$ obeying
\begin{eqnarray}
  \langle \xi(t) \rangle =0, \quad  \langle \xi(t) \xi(t')\rangle = 2
  \eta T \delta(t-t').
\label{xi}
\end{eqnarray}
To account for the quantum nature of the particle and of the bath, we
follow the analysis of quantum Brownian motion by Fisher and
Zwerger\cite{Fisher+85,Zwerger87} who studied the case of a sinusoidal
potential and a dc driving force.

The rectified particle velocity $V$ can be determined from the average
particle coordinate $X(t)$ via
\begin{subequations}
\label{def.V.X}
\begin{eqnarray}
  V &\equiv& \lim_{t \to \infty} \frac 1t X (t) ,
  \label{def.V}
  \\
  X(t) &\equiv& \int dx \ x \ P(t,x) ,
  \label{def.X}
\end{eqnarray}
\end{subequations}
where $P(t,x)$ is the probability distribution for the particle
position at time $t$. This distribution is related to the reduced
density matrix operator $\hrho(t)$ (after the bath degrees of freedom
are traced out) by
\begin{eqnarray}
  P(t,x) =  \langle x | \hrho(t) | x \rangle .
\label{def.P}
\end{eqnarray}
[We use the Dirac notation, where $\langle x^+ | \hrho(t) | x^-
\rangle$ is the density matrix in position representation.]

The dynamics of this density matrix is most conveniently treated in
the Feynman-Vernon path integral
representation.\cite{Feynman+63,Kleinert} The time evolution of the
density matrix from some initial time $t_\rmi$ to a final time
$t_\rmf$ is given by
\begin{subequations}
\label{def.J}
\begin{eqnarray}
  \langle x^+_\rmf | \hrho(t_\rmf) | x^-_\rmf \rangle
  &=&
  \iint d x^+_\rmi d x^-_\rmi
  J(t_\rmf,x_\rmf^+,x_\rmf^-;t_\rmi,x_\rmi^+,x_\rmi^-)
  \nonumber \\ && \times
  \langle x^+_\rmi | \hrho(t_\rmi) | x^-_\rmi \rangle
\label{evol.rho}
\end{eqnarray}
with the kernel
\begin{eqnarray}
  J(t_\rmf,x_\rmf^+,x_\rmf^-;t_\rmi,x_\rmi^+,x_\rmi^-) =
  \int Dx \ Dy \ e^{-S}
\end{eqnarray}
\end{subequations}
being a double path integral over all trajectories $x(t)$ and $y(t)$
with the boundary conditions\cite{note:y}
\begin{eqnarray}
  x(t_{\rmi,\rmf})=\frac 12(x_{\rmi,\rmf}^+ + x_{\rmi,\rmf}^-), \quad
  y(t_{\rmi,\rmf})=\frac 1\hbar (x_{\rmi,\rmf}^+ - x_{\rmi,\rmf}^-).
\end{eqnarray}
The path integral involves the effective action
\begin{subequations}
\label{action}
\begin{eqnarray}
  S&=& S_0 + S_1 ,
  \\
  S_0&=& \frac 12 \iint dt dt' \ y(t) K(t-t') y(t')
  \nonumber \\
  && + i \int dt \ y(t) [m \ddot x(t) + \eta \dot x(t)] ,
  \label{def.S0}
  \\
  S_1&=& i \int dt \Big[\sum_s \frac 1{2s} U(x(t)
  + s y(t))  - y(t) F(t) \Big] ,
\end{eqnarray}
\end{subequations}
with all time integrals running from $t_\rmi$ to $t_\rmf$.  For
notational convenience, the usual contribution $U(x^+(t))-U(x^-(t))$
is written as a sum over the spin-like variable $s=\pm \hbar/2$ which,
however, does not have the meaning of a physical spin.

The effective action (\ref{action}) includes already the average over
the bath degrees of freedom.  This average leads to an integral kernel
$K(t)$ which reads in Fourier representation\cite{note:FT} (we set
$k_{\rm B}=1$)
\begin{eqnarray}
  K(\omega)= {\eta \hbar \omega} \textrm{ coth }  \frac{\hbar
    \omega}{2T} .
\end{eqnarray}
In the classical limit, $K(\omega)=2 \eta T$ reproduces the correlator
(\ref{xi}).  For $T=0$, $K(\w)= \eta \hbar|\w|$ represents a kernel
that is highly nonlocal in time representation.

The model has a large number of parameters: the particle mass $m$ and
the friction coefficient $\eta$, then $\hbar$ and $T$ as a measure of
the strength of quantum and thermal fluctuations.  Further parameters
are implicit in $U(x)$ and $F(t)$.  The potential can be represented
by a Fourier series
\begin{eqnarray}
  U(x)=\sum_q U_q \ e^{i q x}
\end{eqnarray}
with amplitudes $U_q$ for wave vectors $q$.  For periodic potentials
with a period $a$, the wave vectors are
\begin{eqnarray}
  q= n \frac{2\pi}a 
\end{eqnarray}
with integer $n$.

In analogy to $U$,  the ac drive is represented as
\begin{eqnarray}
  F(t)=\sum_\omega F_\omega \ e^{- i\omega t}
\end{eqnarray}
with $F_0=0$ since the force is assumed to be unbiased on time
average.  For a periodic drive with period $t_F$, the frequencies
$\omega$ are integer multiples of the basic frequency $2\pi/t_F$.
Although we assume here periodicity of $U$ and $F$, a generalization
to random $U$ and $F$ is straightforward and will be discussed in the
end of this paper.

\section{Perturbative approach}
\label{sec.pert}

The definition (\ref{def.V.X}) of the velocity has the drawback that
one has to calculate $X(t)$ as the expectation value of the final
position in an ensemble of forward-backward paths of a finite length
$t_\rmf-t_\rmi$.  In order to avoid technical complications related to
boundary effects, we relate the average velocity to an expectation
value at an intermediate time $t$ which can be kept fixed while the
limit $t_\rmi \to - \infty$ and $t_\rmf \to \infty$ is been taken.

Consider the ``partition sum''
\begin{eqnarray}
  Z &=&
  \int d x_\rmf \iint d x^+_\rmi d x^-_\rmi
  J(t_\rmf,x_\rmf,x_\rmf;t_\rmi,x_\rmi^+,x_\rmi^-)
  \nonumber \\ && \times
  \langle x^+_\rmi | \hrho(t_\rmi) | x^-_\rmi \rangle
  \label{def.Z}
\end{eqnarray}
of all forward-backward paths between $t_\rmi$ and $t_\rmf$.  It is
normalized to $Z=1$ since it is the trace of the density matrix at
$t_\rmf$.  We define
\begin{eqnarray}
  V(t) \equiv \langle  \dot x(t) \rangle
\label{def.V.2}
\end{eqnarray}
as expectation value in the ensemble $Z$ of fluctuating paths.  In
this definition, we can take the limit $t_\rmi \to - \infty$ and
$t_\rmf \to \infty$ right away.  In the absence of a nonequilibrium
driving force and due to the presence of dissipation, $V(t)$ would
vanish after an initial relaxation for every possible initial density
matrix $\hrho(t_\rmi)$.  In the presence of the driving force and in
the limit $t_\rmi \to -\infty$, $V(t)$ will be determined uniquely by
$F(t)$ and independently of the initial state.

Although we strictly follow the definition of Fisher and
Zwerger\cite{Fisher+85,Zwerger87} in the path integral formulation of
the problem, we differ in the definition of the average velocity.  We
argue in appendix \ref{app.velo} that, in the long time limit, the
time average of the velocity $V(t)$ coincides with the earlier
definition (\ref{def.V.X}) in combination with (\ref{def.P}) and
(\ref{def.J}).  We find the expectation value $V(t)$ to be a
convenient quantity for the subsequent perturbative evaluation.

\subsection{Perturbative expansion}

To make analytic progress, we consider $F$ as small and calculate the
nonlinear dynamic response of the velocity to the driving force,
\begin{eqnarray}
  V(t) &=& \int dt' \mu_1(t-t') F(t')
  \nonumber \\
  &&+ \frac 1{2!} \iint dt' dt'' \mu_2(t-t',t-t'') F(t') F(t'')
  \nonumber \\
  &&  + O(F^3).
\label{def.mu}
\end{eqnarray}
The mobilities $\mu_m$ can be expressed conveniently as expectation
values in the path ensemble using the partition sum as generating
functional,
\begin{subequations}
\label{mob.exp}
\begin{eqnarray}
  \mu_1(t-t')&=& \left. \frac {\delta V(t)}{\delta F(t')}\right|_{F=0}
  \nonumber \\
  & =& \left. \langle  \dot x(t) iy(t')  \rangle \right|_{F=0} ,
  \\
  \mu_2(t-t',t-t'')&=& \left. \frac {\delta^2 V(t)}
    {\delta F(t') \delta F(t'')}\right|_{F=0}
  \nonumber \\
  & =& \left. \langle  \dot x(t) iy(t') iy(t'') \rangle \right|_{F=0}.
\end{eqnarray}
\end{subequations}
The generalization to higher order mobilities is straightforward.  The
expectation values now refer to the {\em equilibrium} system in the
absence of the driving force.

After Fourier transformation, Eq. (\ref{def.mu}) reads
\begin{eqnarray}
  V_\w &=& \mu_1(\w) F_{\w}
  + \frac 1{2!} \sum_{\w' \w''} \mu_2(\w',\w'') F_{\w'} F_{\w''}
  \delta_{\w,\w'+\w''}
  \nonumber \\
  &&  + O(F^3).
\label{V_0}
\end{eqnarray}
The rectified current is given by the time-average (zero frequency
component) of the velocity,
\begin{eqnarray}
  V_0 = \frac 12  \sum_\w \mu_2(-\w,\w) F_{-\w} F_{\w} + O(F^3).
  \label{def.V_0}
\end{eqnarray}
Since the driving force is unbiased, $F_0=0$, current rectification
cannot be obtained in linear response.  Rather, ratchet effects
require frequency mixing which is present only in nonlinear response.
For weak $F$ the leading ratchet effect will be determined by $\mu_2$.

If the driving force has the symmetry
\begin{eqnarray}
  F(t)=-F(t-t_0)
  \label{F.symm}
\end{eqnarray}
for some time $t_0$ (for example, if $F$ is monochromatic), the
rectified velocity will be invariant under the transformation $F(t)
\to F(t-t_0)=- F(t)$.  Then the contributions to the rectified current
from all mobilities $\mu_m$ with odd $m$ must vanish.

Although the calculation of these mobilities is already much simpler
than a closed calculation of $V(t)$, it still cannot be performed
analytically for general potentials.  Therefore, we employ a second
expansion in $U$, utilizing the weakness of the potential.

The mobilities, which, according to Eqs. (\ref{mob.exp}), are the
equilibrium expectation values $\mu_m=\left. \langle \cO_m \rangle
\right|_{F=0}$ of observables $\cO_m \equiv \dot x(t) iy(t') \cdots
iy(t^{(m)})$, will be calculated perturbatively in the potential using
the expansion\cite{Fisher+85,Zwerger87}
\begin{eqnarray}
  \left. e^{-S_1}\right|_{F=0} = \sum_{n=0}^\infty \frac 1 {n!}
  \left\{ \int dt \sum_{s,q}   \frac{U_{q}}{2 i s}
    e^{iq[x(t) + s y(t)]} \right\}^n .
\end{eqnarray}
We thus can write
\begin{eqnarray}
  \mu_m= \sum_{n=0}^\infty \mu_m^{(n)}
\end{eqnarray}
with
\begin{eqnarray}
  \mu_m^{(n)} &=& \frac 1 {n!} \sum_{\s1,\q1,\cdots,\s{n},\q{n}}
  \prod_{j=1}^n  \frac{U_{\q{j}}}{2 i \s{j}}
  \int dt_1 \cdots \int dt_n
  \nonumber \\
  & &\times
  \Big\langle  \cO_m  \exp \Big({i \sum_{j=1}^n \q{j} [x(t_j) + s_j y(t_j)]}
  \Big) \Big\rangle_0
  \label{mu.m.n}
\end{eqnarray}
where the average $\langle \cdots \rangle_0$ is governed by the
``free'' action $S_0$ defined by Eq. (\ref{def.S0}).  Since $S_0$ is
Gaussian, the averages can be performed straightforwardly using Wick's
theorem.

\subsection{Free theory}

For these averages it is important to know the correlations of the
free theory.  In Fourier representation, one easily finds
\begin{subequations}
\begin{eqnarray}
  \langle x(\w') x(\w'') \rangle_0 &=& C(\w') \delta(\w'+\w'') ,
  \\
  \langle x(\w') iy(\w'') \rangle_0 &=& G(\w') \delta(\w'+\w''),
  \\
  \langle y(\w') y(\w'') \rangle_0 &=&0,
\end{eqnarray}
\end{subequations}
with the response and correlation functions
\begin{subequations}
\label{G.C}
\begin{eqnarray}
  G(\omega)&=&- \frac 1{i\eta \omega + m \omega^2-0^+} ,
  \label{G.w}
  \\
  C(\omega)&=& \frac{K(\omega)}{\eta^2 \omega^2 + m^2 \omega^4} .
\end{eqnarray}
\end{subequations}
To calculate the nonlinear mobilities, one has to use the {\em
  retarded} response function which is in time representation
\begin{eqnarray}
  G(t)&=&\langle x(t) iy(0) \rangle_0 = \frac 1 \eta [1-e^{-\gamma t}]
  \Theta(t)
  \label{G.t}
\end{eqnarray}
with a relaxation rate defined by
\begin{eqnarray}
  \gamma \equiv \frac \eta m.
\end{eqnarray}
Causality of response
\begin{eqnarray}
  G(t)=0 \textrm{ for } t \leq 0
\end{eqnarray}
is reflected by the Heaviside step function $\Theta(t)$ in
(\ref{G.t}).  Note that $\langle x^2(t) \rangle = \infty$ since the
free system is translation invariant and the particle spreads
diffusively (subdiffusively for $T=0$) over the entire space.
Therefore, $C(t)$ is not a well defined quantity.  Instead, the
displacement function
\begin{subequations}
\label{W}
\begin{eqnarray}
  W(t)&\equiv& \frac 12 \langle [x(t) -x(0)]^2 \rangle_0
  \\
  &=& \int \frac{d\omega}{2\pi} [1-\cos(\omega t)] C(\omega)
\end{eqnarray}
captures all information about the particle (sub)diffusion.  This
quantity will play a central role in perturbation theory.
Unfortunately, it can be calculated explicitly only in limiting cases:
\begin{eqnarray}
  W(t)&=& \frac T {\eta \gamma} [\gamma |t| + e^{-\gamma |t|}-1]
  \quad \textrm{ for } \hbar=0,
  \\
  W(t)&\sim& \frac \hbar {\pi \eta} \ln \gamma |t| \quad \textrm{ for } T=0 .
\label{W.qu}
\end{eqnarray}
For semiquantitative purposes,
\begin{eqnarray}
  W(t) \approx \frac T {\eta \gamma} [\gamma |t| + e^{-\gamma |t|}-1]
  + \frac \hbar {2 \pi \eta} \ln [1+(\gamma t)^2]
  \label{W.approx}
\end{eqnarray}
\end{subequations}
is a good interpolation over the whole parameter range.

We conclude this subsection by pointing out some key features of the
response and displacement function.  For $\hbar=0$, $G$ and $W$ are
related through the fluctuation-dissipation relation
\begin{eqnarray}
  T G(t)=  \dot W(t) \Theta(t).
  \label{FDT}
\end{eqnarray}
In the quantum case with $T=0$, $W(t)$ diverges for all $t$ in the
limit $m \to 0$.

\subsection{Characteristic scales}

Before we move on to a further evaluation of the path integral, we
pause for a moment to fix the relevant time, length, and energy scales
of our problem.  From the response function of our problem we can
identify the typical relaxation time
\begin{eqnarray}
  \trel = {\frac 1 \gamma }= \frac m \eta .
\end{eqnarray}
Rewriting the displacement correlation function $W(t)=\ell^2 \hat
W(t/\trel)$ in terms of the dimensionless function $\hat W$ of the
dimensionless argument $t/\trel$, we identify from Eqs.  (\ref{W}) the
diffusion lengths $\ell$ for the thermal and the quantum case,
\begin{eqnarray}
  \lth^2 = \frac {Tm}{\eta^2},
  \quad
  \lqu^2 = \frac \hbar \eta.
\label{def.ell}
\end{eqnarray}
The de-Broglie wave length
\begin{eqnarray}
  \lambda^2 = \frac {2 \pi \hbar^2}{m T} = 2 \pi \frac{\lqu^4}{\lth^2}
\end{eqnarray}
is a related further characteristic scale for the particle in the
absence of dissipation.

Alternatively to Eqs. (\ref{def.ell}), we can associate with thermal
and quantum fluctuations a characteristic energies $E=\eta^2 \ell^2
/m$,
\begin{eqnarray}
  \Eth = T,
  \quad
  \Equ = \hbar \gamma .
\end{eqnarray}
The potential and driving force -- which act as probes to the free
particle -- define the space period $a$, time period $t_F$, and
amplitudes
\begin{eqnarray}
  \cU = 2 |U_q|,
  \quad
  \cF = 2 |F_\w|,
\end{eqnarray}
defined by the lowest harmonic modes $q$ and $\w$.  In the case of
random $U$ or $F$, the periods would be replaced by a correlation
length or time and the amplitudes by variances.

In terms of these scales, a necessary requirement for the validity of
the perturbative approach is that external probes must be weak in
comparison to the internal fluctuations, i.e.,
\begin{eqnarray}
  \cU, a\cF \ll \max (\Eth, \Equ).
  \label{pert.val}
\end{eqnarray}
These scales will also determine the location of the phenomena under
consideration, as we will discuss later on.  However, as we recall by
calculating the linear response mobility, th condition
(\ref{pert.val}) is not sufficient for the validity of perturbative
results.

In the subsequent calculations it is convenient to use dimensionless
quantities.  It is natural to choose as time scale, or $\gamma$ as
frequency scale.  The generic length scale is the potential period
$a$.  The ratios of $\lth^2$ and $\lqu^2$ to $a^2$ provide a natural
measure for the strength of thermal and quantum fluctuations. Hence we
define
\begin{eqnarray}
  \hht \equiv \gamma t,
  \ \
  \hw \equiv \frac \w \gamma,
  \ \
  \hq{j} \equiv a \q{j},
  \ \
  \hhbar \equiv \frac {\hbar}{\eta a^2},
  \ \
  \hT \equiv \frac {Tm}{\eta^2 a^2}.
\label{dimless}
\end{eqnarray}
as dimensionless quantities.

\subsection{Mobilities}

In Eq. (\ref{mu.m.n}), the mobilities $\mu_m^{(n)}$ are determined by
expectation values which can be calculated conveniently from a
generating functional.  We define
\begin{eqnarray}
  \cZ \equiv \Big\langle
  \exp \Big( i \int d\tau [\rho(\tau) x(\tau) + \sigma(\tau) y(\tau) ] \Big)
  \Big\rangle_0
  \label{gen.O}
\end{eqnarray}
as functional of auxiliary fields $\rho(\tau)$ and $\sigma(\tau)$.
$\tau$ is a real time variable and we will be distinguishing it from
$t$ only for bookkeeping purposes.  The averages determining the
mobilities $\mu_m^{(n)}$ can be then represented as functional
derivatives
\begin{eqnarray}
  \Big\langle  \cO_m\exp \Big({i \sum_{j=1}^n \q{j} [x(t_j) + s_j y(t_j)]}
  \Big) \Big\rangle_0 =
  \nonumber \\
  -i \frac d{dt} \frac {\delta^{(m+1)}}{\delta \rho(t)
    \delta \sigma(t') \cdots \delta \sigma(t^{(m)})} \cZ ,
  \label{fun.derive}
\end{eqnarray}
where one has to identify
\begin{subequations}
\label{def.r.s}
\begin{eqnarray}
  \rho(\tau) &=& \sum_{j=1}^n q_j \delta (\tau-t_j),
  \\
  \sigma (\tau) &=& \sum_{j=1}^n q_j s_j \delta(\tau-t_j),
\end{eqnarray}
\end{subequations}
after performing the functional derivatives.  Using the results of the
previous subsection, the generating functional can be expressed as
\begin{eqnarray}
  \cZ &=& \exp \bigg( \iint d\ta1 d\ta2 \Big[-\frac 12 \rho(\ta1)
  C(\ta1-\ta2) \rho(\ta2)
  \nonumber \\
  && +  i \rho(\ta1) G(\ta1-\ta2) \sigma(\ta2) \Big] \bigg) .
\end{eqnarray}
As mentioned previously, $C(t)$ is divergent. This implies that
$\cZ=0$ if $\int d\tau \rho(\tau) = \sum_j q_j \neq 0$.  Therefore,
$\cZ$ can be nonvanishing only if the ``momentum conservation''
$\sum_j q_j =0$ is satisfied.  In this case one may rewrite
\begin{eqnarray}
  \cZ &=& \exp \bigg( \iint d\ta1 d\ta2 \Big[\frac 12\rho(\ta1)
  W(\ta1-\ta2) \rho(\ta2)
  \nonumber \\
  && +  i \rho(\ta1) G(\ta1-\ta2) \sigma(\ta2) \Big] \bigg) \delq.
  \label{res.cZ}
\end{eqnarray}
Hereby, we introduce the abbreviation $\delq \equiv \delta_{\sum
  q_j,0}$.  The subsequent calculations of the mobilities are based on
this generating functional.  For later convenience, we combine Eqs.
(\ref{mu.m.n}), (\ref{fun.derive}), and (\ref{res.cZ}) to our master
formula
\begin{widetext}
\begin{eqnarray}
  \mu_m^{(n)}(t-t',t-t'',\cdots,t-t^{(n)}) &=& -i \frac d{dt}
  \frac 1 {n!} \sum_{\s1,\q1,\cdots,\s{n},\q{n}} \delq
  \prod_{j=1}^n  \frac{U_{\q{j}}}{2 i \s{j}}
  \int dt_1 \cdots \int dt_n
  \frac {\delta^{(m+1)}}{\delta \rho(t)
    \delta \sigma(t') \cdots \delta \sigma(t^{(m)})}
  \nonumber \\
  && \times \exp \bigg( \iint d\ta1 d\ta2 \Big[\frac 12\rho(\ta1)
  W(\ta1-\ta2) \rho(\ta2)
  +  i \rho(\ta1) G(\ta1-\ta2) \sigma(\ta2) \Big] \bigg) .
\label{master}
\end{eqnarray}
\end{widetext}
Thereby, the substitution (\ref{def.r.s}) has to be made after all
functional derivatives are taken.  Momentum conservation implies that
all mobilities vanish for no $n=1$.  For $n=2$ and even $m$ the
mobilities vanish since the contributions to the sum in the right
hand side of Eq. (\ref{master}) are odd in $\{q\}$.  We have noted
already above, that no contribution to the rectified current can arise
from $\mu_m$ with odd $m$ and arbitrary $n$ if the driving force obeys
the symmetry (\ref{F.symm}). In this case, up to fifth order in $F$
and $U$, the only contribution comes from $\mu_2^\3$.

After we have determined the generating functional $\cZ$ for the
mobilities in the previous section, we now turn to the evaluation of
the lowest order mobilities of interest.

\section{Linear mobility $\mu_1$}
\label{sec.mu1}

Although we do not expect ratchet effects from linear response, it is
instructive to calculate $\mu_1$ in order $U^2$ to verify that the
present calculation of the mobility reproduces that results of Fisher
and Zwerger\cite{Fisher+85,Zwerger87} for static $F$ and sinusoidal
$U$ (i.e., for this purpose we include the amplitude $F_0$ in our
consideration).

\subsection{Leading orders}

To zeroth order in $U$, it is obvious that
\begin{eqnarray}
  \mu_1^{(0)}(t-t')=\dot G(t-t') .
\end{eqnarray}
To first order,
\begin{eqnarray}
  \mu_1^{(1)}(t-t')=0
\end{eqnarray}
since the momentum conservation mentioned above cannot be satisfied
(strictly speaking, it is satisfied for the mode $q=0$ which, however,
does not enter the dynamics).

To second order, a straightforward calculation (see appendix
\ref{app.mu1}) leads to
\begin{eqnarray}
  \mu_1^{(2)}(\omega)&=& i\omega G^2(\omega) \sum_q q^2 |U_q|^2
  \Delta B_{-q,q}^{(2)}(\omega)
  \label{mu.1.2}
\end{eqnarray}
with
\begin{subequations}
\begin{eqnarray}
  \Delta B_{-q,q}^{(2)}(\omega)&\equiv&
  B_{-q,q}^{(2)}(\omega=0)-B_{-q,q}^{(2)}(\omega),
  \\
  B_{\q1,\q2}^{(2)}(t)&=&
  \frac 2 \hbar \sin[\frac \hbar 2 \q1^2 G(t)] e^{- \q1^2 W(t)}
 \delta_{\q1+\q2,0} .
\end{eqnarray}
\end{subequations}
In the last expression, the sine has a nonunique sign for
\begin{eqnarray}
  \pi <  \frac \hbar 2  q^2 G(\infty) = \frac {\hbar q^2}{2 \eta} 
\label{osc}
\end{eqnarray}
reflecting quantum interferences of particle trajectories.

We briefly discuss interesting limiting cases of $\mu_1^\2$ for which
we will also examine the ratchet effect later on.

\subsection{Classical limit}

In a classical limit, $\hbar \to 0$, both the overdamped and
underdamped cases are understood fairly
well.\cite{Ambegaokar+69,Ambegaokar+69:err,Ivanchenko+69,Nozieres+79}
In the present perturbative approach, the fluctuation-dissipation
theorem (\ref{FDT}) allows for the simplification
\begin{eqnarray}
  B_{\q1,\q2}^{(2)}(t)=  - \frac 1T \frac d{dt}  e^{- \q1^2 W(t)}
\Theta (t) \delta_{\q1+\q2,0} .
\end{eqnarray}
In this case, the Fourier transformation can be performed
analytically,
\begin{subequations}
\begin{eqnarray}
  \Delta B^{(2)}_{-q,q}(\omega)
  &=& - \frac{i \omega}{T} \int_0^\infty dt\
  e^{i \omega t} e^{- q^2 W(t)}
  \\
  &=& -\frac{i \hw}{T} e^{\hnu_q}
  \hnu_q^{-(\hnu_q- i \hw)} \gamma(\hnu_q-i\hw,\hnu_q),
  \label{B2.class} 
\end{eqnarray}
\end{subequations}
with $\gamma(\cdot,\cdot)$ the incomplete gamma function (to be
distinguished from the parameter $\gamma$).  We introduced the
dimensionless frequency
\begin{eqnarray}
  \hnu_q \equiv \lth^2 q^2 =    \frac{Tm q^2}{\eta^2}
  \label{def.nu}
\end{eqnarray}
related to the thermal diffusion time over a distance $1/q$ via
$W( \trel/\hnu_q)=q^{-2}$.  The insertion of expression
(\ref{B2.class}) into Eq. (\ref{mu.1.2}) yields an explicit analytic
expression for the classical linear response mobility at finite
frequencies,
\begin{eqnarray}
  \mu_1^{(2)}(\omega)&=& - \frac 1 \eta \frac 1{(1-i\hw)^2}
  \nonumber \\
  & \times& \sum_q \frac{|U_q|^2}{T^2} e^{\hnu_q}
  \hnu_q^{1-(\hnu_q- i \hw)} \gamma(\hnu_q-i\hw,\hnu_q)
\end{eqnarray}
which reproduces Eq. (4.11) of Ref. \onlinecite{Fisher+85} for the
special case of sinusoidal $U$ and $\w=0$.

In the massless (overdamped) limit $m \to 0$, where $B$ can be easily
Fourier transformed, this simplifies to
\begin{eqnarray}
  \mu_1^{(2)}(\omega)&=&- \frac 1{\eta}  \frac{\cU^2}{T^2}
  \hmu_1^\2 \left(  \frac {\eta a^2 \w} T  \right)
\end{eqnarray}
with a dimensionless scaling function
\begin{eqnarray}
  \hmu_1^\2(z) &=& \sum_q \frac{|U_q|^2}{\cU^2}  \frac
  {\hq{}^2} {\hq{}^2 - i z} .
\end{eqnarray}
Thus, for $m=0$, the corrections to mobility decay proportional to
$T^{-2}$ at high temperatures.

\subsection{Adiabatic limit}

The adiabatic limit $\omega \to 0$ simplifies the calculation of
$\Delta B_{-q,q}^{(2)}(\omega)$, resulting in
\begin{eqnarray}
  \mu_1^{(2)}(\omega=0) &=& - \frac{2}{\eta^2 \hbar} \sum_q |U_q|^2 q^2
  \nonumber \\ && \times
  \int_0^\infty dt \ t \ e^{- q^2 W(t)} \sin[\frac \hbar 2 q^2 G(t)]
\end{eqnarray}
which agrees with the linear response limiting case (4.18) of
Ref. \onlinecite{Fisher+85}.

At $T=0$, the particle shows a remarkable localization transition due
to the dissipative coupling.\cite{Schmid83,Fisher+85} For strong
coupling, the particle is localized in an arbitrarily weak potential,
whereas it remains mobile for weak damping.  This transition is
reflected by the divergence of the mobility correction
$\mu_1^{(2)}(\omega=0)$ due to a divergence of the time integral at
large $t$.  From the logarithmic asymptotics (\ref{W.qu}) of $W(t)$
one can identify the location of the transition at $\alpha=1$ with
\begin{eqnarray}
  \alpha \equiv \frac{\eta a^2}{2\pi\hbar} =
  \frac 1{2\pi \hhbar} .
\label{def.alpha}
\end{eqnarray}
Note that for $\alpha <1$ the inequality (\ref{osc}) is fulfilled for
all wave vectors $q$, i.e. quantum interference effects suppress the
contribution to $\mu_1^\2$.

In the strong damping regime, the divergence of $\mu_1^\2(\omega=0)$
signals the break-down of perturbation theory.  Thus, at $T=0$ the
condition $\alpha \ll 1$ should be added to the condition
(\ref{pert.val}).  This condition may be regarded also as a condition
for the period of the potential (with localization for $a^2 \leq 2 \pi
\lqu^2$).

In the perturbatively accessible regime of $\alpha <1$, Fisher and
Zwerger\cite{Fisher+85} pointed out the interesting fact that the
mobility is a {\em nonmonotoneous} function of temperature.  At zero
temperature, the particle has its free mobility.  Weak thermal
fluctuations ($T<T^*$) first reduce mobility (thermally resisted
quantum tunneling), whereas strong thermal fluctuations increase
mobility back to its free value (thermally assisted hopping).  The
crossover occurs for $\alpha \ll1$ at the temperature
\begin{eqnarray}
  T^* \simeq \frac {\pi^2 \hbar^2}{3 m a^2}
\end{eqnarray}
at which the de-Broglie wave length is comparable to the potential
period, $\lambda \simeq a$.

Before we continue to enter new territory, we wish to conclude this
subsection by stressing that our approach successfully reproduces
previous linear response results for $\w=0$ and already provides
additional insight into the frequency dependence.

\section{Nonlinear mobility $\mu_2$}
\label{sec.mu2}

The generating functional formalism presented above provides an
efficient tool to calculate also higher order mobilities.  Here, we
focus on the lowest order of $\mu_2$ in $U$ contributing to current
rectification.

\subsection{Leading orders}

To zeroth order,
\begin{eqnarray}
  \mu_2^{(0)}(t-t',t-t'')=\langle \dot x(t) iy(t') iy(t'') \rangle_0 = 0
\end{eqnarray}
vanishes since $S_0$ is invariant under the reflection $\{x,y\} \to
\{-x,-y\}$.  To first order,
\begin{eqnarray}
  \mu_2^{(1)}(t-t',t-t'')=0
\end{eqnarray}
vanishes again because momentum conservation cannot be satisfied.  To
second order,
\begin{eqnarray}
  \mu_2^{(2)}(t-t',t-t'')=0
\end{eqnarray}
according to the general statements following Eq. (\ref{master}).

The general third order contribution $\mu_2^\3(t-t',t-t'')$ is given
by expression (\ref{mu.2}) calculated in appendix \ref{app.mu2}.
Since this expression is somewhat clumsy and since we are interested
only in ratchet effects, we can restrict our considerations to
\begin{widetext}
\begin{eqnarray}
  \mu_2^\3(-\omega,\omega)
  &=& - \frac i{\eta} \frac 1{\eta^2 \w^2 + m^2 \w^4} \sum_{\q1\q2\q3}
  \U1 \U2 \U3 \q1
  \nonumber
  \\
  && \times
  \{ \q1 \q2 [2 \B(0,0) - \B(-\w,0) - \B(\w,0)]
  \nonumber
  \\
  &&
  + \q1 \q3 [2 \B(0,0) - \B(-\w,-\w) - \B(\w,\w)]
  \nonumber
  \\
  &&
  + \q2 \q3 [2 \B(0,0) - \B(0,-\w) - \B(0,\w)]
  \}
\label{mu.2.3}
\end{eqnarray}
with
\begin{eqnarray}
  \B (t_1-t_2,t_2-t_3)
  &=&   \frac 2\hbar \sin[\frac\hbar2 \q1 G_{12} \q2]
  \frac 2\hbar \sin[\frac\hbar2 (\q1 G_{13} \q3 + \q2 G_{23}\q3 )]
  \nonumber
  \\
  && \times
  \exp (\q1 W_{12} \q2 + \q2 W_{23} \q3 + \q1 W_{13} \q3)
\delta_{\q1+\q2+\q3,0} \Theta_{23}.
\label{def.B3}
\end{eqnarray}
\end{widetext}
Note that $\B$ is an implicit function of $\{q\}$ and invariant under
$\{q\} \to - \{q\}$.  Consequently, $\mu_2^{(3)}(-\w,\w)$ changes sign
under a reflection $U_q \to U_{-q}$ which implies that the rectified
velocity (\ref{def.V_0}) vanishes for even potentials as it should.
Examining the contribution from a set of wave vectors $\{q\}$ and its
reflected set $-\{q\}$, one can recognize that $\mu_2^{(3)}(-\w,\w)$
is real and that it depends only on
\begin{eqnarray}
 \hU \equiv \textrm{Im } \frac{\U1 \U2 \U3}{\cU^3}.
\end{eqnarray}
(In order to obtain simple analytic expressions below, we find it most
convenient to use momentum conservation to eliminate the sum over
$\q2$.)  In Eq.  (\ref{mu.2.3}), the sum over the momenta can be
restricted to $\q{j} \neq 0$ (in addition to momentum conservation)
since $\B$ vanishes otherwise (later on, these restrictions are
referred to by $\sum'$).  Physically, it is clear that a constant
shift of the potential cannot enter the dynamics of the particle.

Eq.  (\ref{mu.2.3}) is our main result in general form.  Further
analytic progress is hampered by the absence of an analytic expression
for $W(t)$.  Nevertheless, further analytic progress is possible in
various limiting cases.

Using the dimensionless quantities defined in Eqs. (\ref{dimless}) we
may reexpress Eq. (\ref{mu.2.3}) as
\begin{eqnarray}
  \mu_2^\3(-\omega,\omega) = \frac {\cU^3}{\eta^3 a^3 \hbar^2 \gamma^4}
    \hmu_2^\3 (\hhbar,\hT,\hw)
\end{eqnarray}
with $\hmu$ being a dimensionless function of dimensionless arguments.
For a monochromatic driving force, the rectified velocity is
\begin{eqnarray}
  V_0= \frac {\cU^3 \cF^2}{4 \eta^3 a^3 \hbar^2 \gamma^4}
  \hmu_2^\3 (\hhbar,\hT,\hw)
\end{eqnarray}
to leading order according to Eq. (\ref{V_0}).

\subsection{Limit $\hbar \to 0$ and $m \to 0$}
\label{sec.class}

We start with the examination of the classical limit.  To further
simplify the analysis and to perform a comparison of our perturbative
results with previous approaches, we consider the overdamped limit
with $m=0$. In this case, the memory kernel $\B$ simplifies
considerably to
\begin{eqnarray}
  \B (\w_1,\w_2) = - \delq  \frac {\q1 \q2 \q3^2 } {\eta^2}
  \frac 1{\nuq1 - i \w_1} \frac 1{\nuq3 - i \w_2}
  \label{B3.class}
\end{eqnarray}
with the characteristic frequencies
\begin{eqnarray}
  \nu_q \equiv T q^2/\eta.
\end{eqnarray}
This frequency corresponds to the time $\nu_q^{-1}$ a classical
particle needs to diffuse over a distance $q^{-1}$.  Insertion of
Eq. (\ref{B3.class}) into Eq. (\ref{mu.2.3}) leads to
\begin{eqnarray}
  \hmu_2^\3 (\hhbar,\hT,\hw) \to \hhbar^2 \hT^{-4} \hmucl(\hw/\hT),
\end{eqnarray}
i.e.,
\begin{eqnarray}
  \mu_2^{(3)}(-\omega,\omega) =
  \frac {a \cU^3}{\eta T^4} \hmucl\left( \frac{ \eta a^2 \w}T
  \right)
  \label{res.class}
\end{eqnarray}
with a reduced scaling function
\begin{eqnarray}
  \hmucl(z) &\equiv&
  - 2 \sum_{\q1\q2\q3}' \hU  (\hq1 + \hq3 )
  \nonumber \\ && \times
  \left\{\frac {2\hq1^2} {\hq1^4 + z^2}
    - \frac{\hq1\hq3(\hq1^2 \hq3^2  - z^2)}
    {(\hq1^4 + z^2)(\hq3^4 + z^2)}  \right\} .
  \label{class.scal}
\end{eqnarray}
The primed sum is restricted to momenta satisfying momentum
conservation and $\q{j} \neq 0$.  It is interesting to realize that
the function $\hmucl(z)$ is uniquely determined by the shape of the
potential.  If current reversals exist in the limit under
consideration, they correspond to oscillatory behavior of this
function.

The scaling function $\hmucl(z)$ becomes simple in additional
limiting cases.  In deriving these limits from Eq. (\ref{class.scal}),
one has to make use of momentum conservation and of permutations of
momentum labels.  For $z \to \infty$
\begin{subequations}
\label{class.scal.lim}
\begin{eqnarray}
  \hmucl(z) &\to&
  \frac{4}{z^4} \sum_{\q1\q2\q3}' \hU  \hq1^3 \hq2^3 \hq3
  \nonumber \\
  &&= \frac{4a^6}{z^4 \cU^3} \int_0^a dx \ [U'''(x)]^2 U'(x).
\end{eqnarray}
The terms of order $z^{-2}$ cancel each other.  Thus, we easily
retrieved the result obtained previously in Ref.
\onlinecite{Reimann00:lnp}.

In the opposite limit $z \to 0$,
\begin{eqnarray}
  \hmucl(z) &\to& 2 \sum_{\q1\q2\q3}' \hU
  \frac{1}{\hq1}
  \nonumber \\
  &&= \frac 2{a^2 \cU^3} \int_0^a dx \ U^2(x) \delta \Upsilon(x)
  \label{class.w.eq.0}
\end{eqnarray}
\end{subequations}
with a potential integral
\begin{subequations}
\begin{eqnarray}
  \Upsilon(x) &\equiv& \int_0^x dy \ U(y),
  \\
  \delta \Upsilon(x) &\equiv& \Upsilon(x)-
  \frac 1a \int_0^a dy \ \Upsilon(y).
\end{eqnarray}
\end{subequations}
In deriving Eq. (\ref{class.w.eq.0}) we have assumed $U_0=0$,
otherwise additional subtraction terms should be added.

This scaling behavior (\ref{class.scal.lim}) implies the asymptotic
behaviors of the rectified velocity (\ref{def.V_0})
\begin{subequations}
\begin{eqnarray}
  V_0 &\propto& T^0 \w^{-4} \textrm { for } \w \to \infty,
  \\
  V_0 &\propto& \w^0 T^{-4} \textrm { for } T \to \infty.
\label{V0.hT.od}
\end{eqnarray}
\end{subequations}
The apparent divergence of $V_0$ for $T \to 0$ is an artifact of
leaving the range of validity of our perturbative approach.
Analogously, the apparent divergence of $V_0$ for $\eta \to 0$ is due
to the assumption of overdamped dynamics.  Nevertheless, these
divergences may be interpreted as indications that ratchet effects
are particularly strong at low $T$ and in the underdamped case. This
situation will be examined later on.

\begin{figure}
  \includegraphics[angle=270,width=0.9\linewidth]{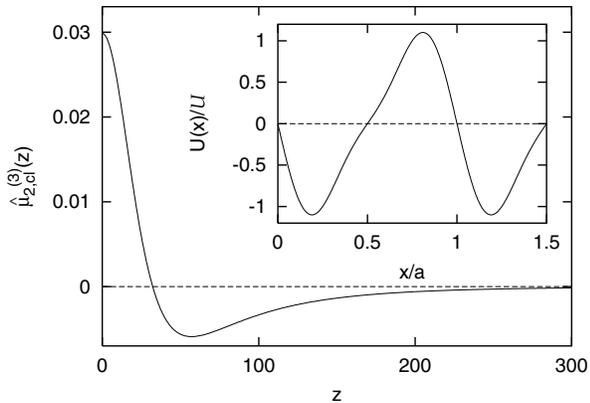}
  \caption{
    Scaling function $\hmucl(z)$ for the potential (\ref{U.ex})
    shown in the inset.}
\label{fig_scal1}
\end{figure}

Before we move on to other limiting cases, we show that the current
reversal phenomenon is captured by our perturbative approach.  We find
it also instructive to perform an explicit quantitative comparison of
our results with the exact results in Ref. \onlinecite{Bartussek+94}.
For this comparison, we evaluate Eqs.  (\ref{def.V_0}) with
(\ref{res.class}) for
\begin{eqnarray}
  U(x)=-\cU\left[\sin(2\pi \frac xa) + \frac 14 \sin(4\pi \frac xa)\right],
  \label{U.ex}
\end{eqnarray}
cf. the inset of Fig. \ref{fig_scal1}.  The corresponding scaling
function (\ref{class.scal}) is shown in Fig. \ref{fig_scal1}.  Since
it has one zero, we expect one current reversal.

\begin{figure}
\includegraphics[angle=270,width=0.9\linewidth]{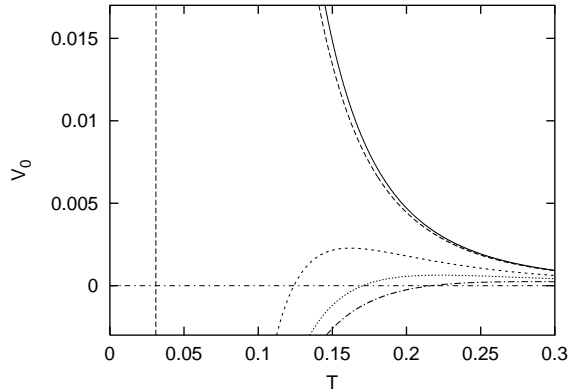}
\caption{$V_0(T)$ for $\w=0.01$ (bold line), $\w=1$ (long dashes), $\w=4$
  (short dashes), $\w=5.5$ (dotted), and $\w=7$ (dash-dotted) for
  comparison with Fig. 1a in Ref. \protect\onlinecite{Bartussek+94}.
  The vertical line represents the vicinity of the current reversal
  for $\w=1$.}
\label{fig_vT_class}
\end{figure}

We explicitly compare our perturbative result for the rectified
velocity as a function of temperature -- displayed in Fig.
\ref{fig_vT_class} -- with the exact solution displayed in Fig. 1a of
Ref.  \onlinecite{Bartussek+94}.  Thereby, length, energy and time
scales are fixed by the choices $a=1$, $\cU=\frac 1{2\pi}$, and
$\eta=1$.  The monochromatic driving force is $F(t)=\cF \sin(\w t)$
with amplitude $\cF=0.5$.  From the shape of the scaling function it
is clear that we find a current reversal with varying temperature for
every $\w>0$ and also a current reversal with varying frequency for
every finite $T$.  The {\em quantitative} agreement is good for $T
\gtrsim \cU$, where the perturbation theory in $U$ is justified.

\begin{figure}
  \includegraphics[angle=270,width=0.9\linewidth]{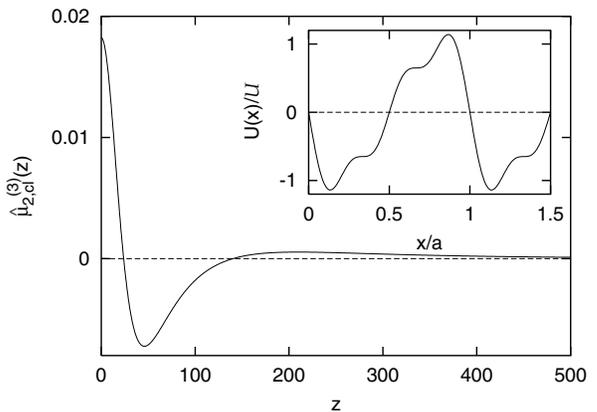}
  \caption{
    Scaling function $\hmucl(z)$ for the potential (\ref{U.ex2})
    shown in the inset.}
\label{fig_scal2}
\end{figure}

The classical limit is not restricted to single current reversals.  It
is likely that an arbitrary number of current reversals can be
obtained by suitably tailored potential.  We have found, for example,
that it is sufficient to add one more harmonic to obtain a second
current reversal.  Specifically, the potential
\begin{eqnarray}
  U(x)=-\cU\left[\sin(2\pi \frac xa) + \frac 14 \sin(4\pi \frac xa)
  + \frac 14 \sin(6\pi \frac xa)\right]
  \label{U.ex2}
\end{eqnarray}
leads to the scaling function shown in Fig. \ref{fig_scal2} with two
zeroes, i.e., two current reversals.

\subsection{Limit $T \to \infty$ for $m>0$}

It is interesting to examine the high-temperature limit since there
are significant differences between the cases $m=0$ and $m>0$.  For $T
\to \infty$, the exponential factor in Eq. (\ref{def.B3}) strongly
suppresses $\B$ and thus $\mu_2^\3$.  In this limit, we find the asymptotic
behavior of the mobility (see appendix \ref{app.hT})
\begin{subequations}
\label{mu.hT}
\begin{eqnarray}
  \hmu_2^\3(\hhbar,\hT,\hw) \to \frac {\hhbar^2 \hT^{-17/6}}{1+\hw^2} \hmuhT ,
\end{eqnarray}
or
\begin{eqnarray}
  \mu_2^\3(-\omega,\omega)
  =  \frac 1 \eta \frac {a \cU^3}{(\eta a^2 \gamma)^4}
  \left( \frac{\eta a^2 \gamma}{T} \right)^{17/6}
  \frac{\hmuhT}{1+\w^2/\gamma^2}
\end{eqnarray}
with
\begin{eqnarray}
  \hmuhT &\equiv& 2 \sqrt{2 \pi}
  \Gamma(\frac 13) \sum_{\q1>0,\q3>0} \hU \hq1^2
  \nonumber \\ && \times
  \left( \frac{(\hq1+\hq3)^2 \hq3}{\hq1^5} \right)^{1/3}
  \left( \frac{3^{1/3} \hq1^2}{\hq1^2+\hq3^2} - \frac 16 \right) .
\end{eqnarray}
\end{subequations}
The constant $\hmuhT$ is uniquely determined by the shape of the
potential.  In the high temperature limit, the rectified current is
much stronger for massive particles than for massless cases, since
$\mu_2^\3 \sim T^{-17/6}$ for $m>0$ whereas $\mu_2^\3 \sim T^{-4}$ for
$m=0$, cf. Eq. (\ref{V0.hT.od}).  This observation is consistent with
the mass dependence in Eq.  (\ref{mu.hT}), $\mu_2^\3 \sim m^{7/6}
T^{-17/6}$ for $m \to 0$, which signals that in the limit $m \to 0$,
$\mu_2^\3$ should decay with a higher power of temperature.  Thus,
{\em inertial terms are crucial} at high temperatures even for large
friction where the relaxation of the particle in the minima of the
potential is overdamped.

\subsection{Limit $\w \to \infty$}

For large frequencies, Eq. (\ref{mu.2.3}) simplifies to
\begin{subequations}
\begin{eqnarray}
  \hmu_2^\3(\hhbar,\hT,\hw) \to  \frac1{\hw^2(1+\hw^2)} \hmuhf(\hhbar,\hT)
\end{eqnarray}
or
\begin{eqnarray}
  \mu_2^\3(-\omega,\omega)
  =  \frac {a \cU^3}{\eta (\eta a^2 \w)^2 \hbar^2
    (\gamma^2+  \w^2) }
  \hmuhf (\hhbar,\hT)
\end{eqnarray}
with
\begin{eqnarray}
   \hmuhf(\hhbar,\hT)
  &\equiv&  - \sum_{\q1\q2\q3} \hU a^3 \q1 \{\q1^2 + \q1 \q3 + \q3^2\}
  \nonumber
  \\
  && \times \hbar^2 \gamma^2 \B(0,0).
\label{mu.hf}
\end{eqnarray}
\end{subequations}
[For the discussion of this limit, $\B(0,0) \equiv \B(\w=0,\w=0)$.]
$\hmuhf$ is a function of the potential shape and of to parameters
measuring the strength of quantum and thermal fluctuations.  In the
special case $\hbar=m=0$ we found in Sec. \ref{sec.class} a momentum
dependence $\B(0,0) \propto \q2/\q1$ which led to a cancellation in
the sum over momenta in expression (\ref{mu.hf}).  Using the
fluctuation-dissipation relation (\ref{FDT}), one can easily show that
$\B(0,0)$ is independent of $m$ for $\hbar=0$.  Thus, this
cancellation persists as long as $\hbar=0$, i.e., $\hmuhf(0,z)=0$.
For $m>0$ this implies a decay $\mu_2^\3(-\omega,\omega) \propto
\w^{-6}$.  However, such a cancellation can no longer be expected for
$\hbar>0$. In this case, one finds again $\mu_2^\3(-\omega,\omega)
\propto \w^{-4}$ at large frequencies.

\subsection{Limit $\w \to 0$}

A further limit of interest is the adiabatic limit for the quantum
particle.  This limit was also studied in the
past\cite{Reimann+97,Yukawa+97,Tatara+98} and has revealed that
additional current reversals may arise from the competition of quantum
and thermal fluctuations.

\begin{figure}
\includegraphics[angle=270,width=0.9\linewidth]{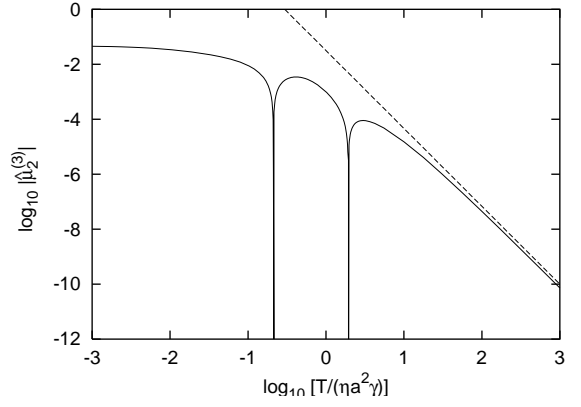}
\caption{
  Double logarithmic plot of the rectified velocity [the dimensionless
  quantity $\hmuad$ given in Eq. (\ref{mu.2.3.ad})] versus temperature
  in the adiabatic limit for the potential (\ref{U.ex}) in the
  underdamped case $\alpha=\frac 12$ (bold line). The dashed line is a
  guide to the eye representing the behavior $\propto T^{-17/6}$ of
  the high-temperature limit.}
\label{fig_scanT}
\end{figure}

For $\w \to 0$, Eq.  (\ref{mu.2.3}) reduces to
\begin{subequations}
\label{mu.2.3.ad}
\begin{eqnarray}
  \hmu_2^\3(\hhbar,\hT,0)
  &\equiv&
  - \sum_{\q1\q2\q3} \hU \iint_0^\infty d\hat t'  d\hat t''
  \nonumber \\  &&  \times
  \hq1 (\hq1 \hat t'-\hq3 \hat t'')^2 \hB(\hat t',\hat t'')
\end{eqnarray}
with
\begin{eqnarray}
  \hB(\hht',\hht'')  \equiv  \hbar^2 \B(\hht'/\gamma,\hht''/\gamma)
\end{eqnarray}
\end{subequations}
We have calculated $\hmuad$ numerically for $\hhbar=\frac 1 \pi$ [i.e.
$\alpha = \eta a^2/(2 \pi \hbar) = \frac 12$ corresponding to the
delocalized case, cf. Eq.  (\ref{def.alpha})] as a function of
temperature (cf. Fig.  \ref{fig_scanT}) for the potential
(\ref{U.ex}).  The two poles of the double-logarithmic plot in Fig.
\ref{fig_scanT} represent current reversals.  At high temperatures,
the relation $\hmu_2^\3 \propto T^{-17/6}$ is recovered (dashed line).
At zero temperature, a finite current is generated by quantum
fluctuations.

\section{Conclusions}
\label{sec.disc}

We have developed a perturbative approach for quantum ratchets, which
captures current rectification and reversals of the current direction.
Our main results are the analytic expression (\ref{mu.2.3}) for the
leading nonlinear mobility and its evaluation for various limiting
cases.  In particular, the high-temperature limit for massive
particles has revealed the relevance of inertial terms even for strong
damping. Since the rectified current decays like $V_0 \propto T^{-4}$
for massless particles whereas it decays like $V_0 \propto T^{-17/6}$
for massive particles, inertial effect can lead to a substantial {\em
  enhancement} of ratchet effects.  On the other hand, in the
high-frequency limit, the quantum nature of the particle is important.
While $V_0 \propto \w^{-6}$ for massive classical particles, quantum
fluctuations also {\em enhance} the rectified currant, leading to $V_0
\propto \w^{-4}$.

While our perturbative approach is limited to weak potentials and
driving forces, it has the advantage that it can be easily generalized
to higher dimensions.  Therefore, applications for example to
asymmetric antidot arrays\cite{Lorke+98} become possible.
Furthermore, a generalization to random ratchet potentials is obvious.
Thereby one could describe the case of asymmetric potential wells with
random positions.  This generalization can be achieved if one allows
for continuous wave vectors $q$ of the potential and simply replaces
$\U1 \U2 \U3$ by its average in the nonlinear mobility (\ref{mu.2.3}).
An extension of this perturbative approach from single quantum
particles to electron gases is under current investigation by the
authors.

\begin{acknowledgments}
  We thank Yu. Nazarov and W. Zwerger for pointing out to us
  references related to our work.  This work was supported by the U.S.
  Department of Energy, Office of Science through contract No.
  W-31-109-ENG-38
\end{acknowledgments}

\appendix

\section{Velocity}
\label{app.velo}

In this appendix we show that the average velocity can be calculated
from the definition (\ref{def.V.2}).  It is this definition in which
we deviate from the approach of Fisher and
Zwerger.\cite{Fisher+85,Zwerger87}

The original definition (\ref{def.V}), in a more explicit form, is
based on the average distance the particle travels in a large time
interval,
\begin{eqnarray}
  V &\equiv&  \frac {X (t_\rmf) - X(t_\rmi)}{t_\rmf-t_\rmi}
\end{eqnarray}
in the limit $t_\rmf \to \infty$ and $t_\rmi \to -\infty$.  Thereby the
position expectation value (\ref{def.X}) and the time evolution
(\ref{evol.rho}) of the density matrix.  On the other hand, the time
average of $V(t)$, Eq. (\ref{def.V.2}),  can be written as
\begin{eqnarray}
  \overline{V(t)} &\equiv& \frac {\tX (t_\rmf) - \tX(t_\rmi)}{t_\rmf-t_\rmi}
\end{eqnarray}
with
\begin{subequations}
\begin{eqnarray}
  \tX(t) &\equiv& \int dx \ x \tP(t,x),
  \\
  \tP(t,x) &\equiv& \langle \delta(x-x(t)) \rangle.
\end{eqnarray}
\end{subequations}
{\em A priori}, $\tP(t,x)$, which is an expectation value in an
ensemble of paths of length $t_\rmf-t_\rmi$, is different from
$P(x,t)$, Eq. (\ref{def.P}), which is an expectation value in an
ensemble of paths of length $t-t_\rmi$.  However, the definitions
coincide for $t=t_\rmf$ and also for $t=t_\rmi$. In the first case the
definitions coincide, in the second case because one can integrate out
the paths (the integral corresponds to $Z$, Eq. (\ref{def.Z}), the
integral is most easily performed for a diagonal initial density
matrix). Thus,
\begin{eqnarray}
  {\tX (t_\rmf) - \tX(t_\rmi)} ={X (t_\rmf) - X(t_\rmi)}
\end{eqnarray}
and $V=\overline{V(t)}$.  If there is a well defined expectation value
$V(t)\equiv \langle \dot x(t) \rangle =\frac d{dt} \tX(t)$ for $t_\rmi
\to - \infty$ and $t_\rmf \to \infty$, it must coincide with
$\overline{V(t)}$ since boundary effects from times near $t_\rmi$ and
$t_\rmf$ should become negligible in this limit.

\section{Calculation of $\mu_1^{(2)}$}
\label{app.mu1}

Here we present intermediate steps of the calculation leading to
Eq. (\ref{mu.1.2}).  In a first step, we need to evaluate
\begin{eqnarray}
 \frac{\delta^2}{\delta \rho(t) \delta \sigma(t')} \cZ &=&
\Big\{ i G(t-t') + \iint d\ta1 d\ta2 [W(t-\ta1) \rho(\ta1)
\nonumber \\
 &+& i G(t-\ta1) \sigma(\ta1) ]  \rho(\ta2) iG(\ta2-t') \Big\} \cZ
\label{d2Z}
\end{eqnarray}
where Eqs. (\ref{def.r.s}) have to be inserted for $n=2$.  Thereby,
\begin{eqnarray}
   \cZ = e^{  \q1  W_{12}\q2  +  \q1 iG_{12} \s2 \q2
   +  \q2 i G_{21}\s1 \q1 } \delta_{\q1+\q2,0}
\end{eqnarray}
where we abbreviate $W_{kl} \equiv W(t_k-t_l)$ etc. and we used
$W(0)=G(0)=0$.  Note that in the last exponential, $G_{12}$ or
$G_{21}$ vanishes for all $t_1$, $t_2$ because of causality.
Inserting Eq. (\ref{d2Z}) into Eqs. (\ref{gen.O}) and (\ref{mu.m.n}),
only the last of the three terms coming from Eq. (\ref{d2Z}) survives
summation over $s_j$ in Eq. (\ref{mu.m.n}).  One obtains
\begin{widetext}
\begin{eqnarray}
  \mu_1^\2(t-t')&=&  - i \sum_{\q1\q2\s1\s2}
\frac {\U1 \U2}{4 \s1 \s2} \iint dt_1dt_2
 \dot G(t-t_1) \s1\q1
[\q1 G(t_1-t') + q_2 G(t_2-t')] \cZ.
\end{eqnarray}
In these remaining terms, the summation over $\s{j}$ yields
\begin{subequations}
\begin{eqnarray}
 \mu_1^\2(t-t') &=& - \sum_{\q1\q2} \U1\U2 \iint dt_1dt_2
\dot G(t-t_1) \q1
[\q1 G(t_1-t') + q_2 G(t_2-t')] B_{\q1\q2}^\2(t_1-t_2),
\\
B_{\q1\q2}^\2(t_1-t_2)&\equiv& \sum_{\s1 \s2} \frac i{4 \s1 \s2} \s1 \cZ =
  \frac 2 \hbar \sin (\frac \hbar 2 \q1^2 G_{12}) \exp(- \q1^2 W_{12})
  \delta_{\q1+\q2,0}.
\end{eqnarray}
\end{subequations}
Fourier transforming this expression leads to Eq. (\ref{mu.1.2}).

\section{Calculation of $\mu_2^\3$}
\label{app.mu2}

Following the same route as for  $\mu_1^\2$, we first calculate
\begin{eqnarray}
  \frac {\delta^3}{\delta \rho(t) \delta \sigma(t') \delta \sigma(t'')} \cZ
  &=&
  \int d\tau \rho(\tau) [i G(t-t') iG(\tau - t'') +
  i G(t-t'')  iG(\tau - t') ]\cZ
  \nonumber \\
  &&+ \int d\tau [W(t-\tau)\rho(\tau) + iG(t-\tau)\sigma(\tau)]
    \nonumber \\
  && \times
  \int d \tau' \rho(\tau') iG(\tau'-t')
  \int d \tau'' \rho(\tau') iG(\tau''-t'') \cZ .
  \label{der.Z.3}
\end{eqnarray}
Eq. (\ref{def.r.s}) leads to
\begin{eqnarray}
  \cZ=e^{\q1 W_{12} \q2 + \q1 W_{13} \q3 + \q2 W_{23} \q3
    + [\q2 iG_{21} + \q3 iG_{31}] \q1\s1
    + [\q1 iG_{12} + \q3 iG_{32}] \q2\s2
    + [\q1 iG_{13} + \q2 iG_{23}] \q3\s3}
  \delq
\end{eqnarray}
with $\delq \equiv \delta_{\q1+\q2+\q3,0}$.  Considering the
right-hand side of Eq. (\ref{der.Z.3}) as a sum of four contributions,
the first three disappear after summation over the $s_j$.  For
example, if $t_1<t_2<t_3$, $\cZ$ is independent of $s_3$ and the
summation over $s_3$ leads to a cancellation.  The remaining fourth
contribution to Eq. (\ref{der.Z.3}) reads explicitly [$\Theta_{23}
\equiv \Theta(t_2-t_3)$]
\begin{eqnarray}
  \mu_2^\3(t-t',t-t'')&=& - \sum_{\q1,\q2,\q3,\s1,\s2,\s3}
  \frac{\U1}{2i\s1} \frac{\U2}{2i\s2}  \frac{\U3}{2i\s3}
  \iiint dt_1 dt_2 dt_3 \ \Theta_{23} \dot G(t-t_1) \s1 \q1
  \nonumber \\
  && \times \{
  \q1 G(t_1-t') \q1 G(t_1-t'') +
  \q2 G(t_2-t') \q2 G(t_2-t'') +
  \q3 G(t_3-t') \q3 G(t_3-t'')
  \nonumber \\
  && + \q1 \q2 [G(t_1-t') G(t_2-t'') + G(t_1-t'') G(t_2-t')]
  \nonumber \\
  && + \q1 \q3 [G(t_1-t') G(t_3-t'') + G(t_1-t'') G(t_3-t')]
  \nonumber \\
  && + \q2 \q3 [G(t_2-t') G(t_3-t'') + G(t_2-t'') G(t_3-t')]
  \} \cZ
\end{eqnarray}
where we used permutation symmetries among indices $j$ which allow us
to restrict the time integrals to $t_2>t_3$.  Then, summation over
$\{s\}$ leads to
\begin{eqnarray}
  B_{\q1\q2\q3}^\3(t_1-t_2,t_2-t_3) &\equiv& \Theta_{23} \sum_{\s1\s2\s3}
  \frac 1{2\s1} \frac  1{2\s2} \frac 1{2\s3} \s1 i^2 \cZ
  \\
  &=&   \Theta_{23} \frac 4 {\hbar^2} \sin[\frac\hbar2 \q1 G_{12} \q2]
  \sin[\frac\hbar2 (\q1 G_{13} \q3 + \q2 G_{23}\q3 )]
  e^{\q1 W_{12}\q2 + \q2  W_{23}\q3 + \q1 W_{13} \q3} \delq .
\end{eqnarray}
Note that $\B \neq 0 $ only for $t_1>t_2>t_3$.  In terms of $\B
\equiv B^\3_{\q1\q2\q3}$ we obtain
\begin{subequations}
\label{mu.2}
\begin{eqnarray}
  \mu_2^\3(t-t',t-t'')&=& i \sum_{\q1,\q2,\q3}
  \U1\U2\U3
  \iiint dt_1 dt_2 dt_3 \ \dot G(t-t_1) \q1
  \nonumber \\
  && \times \{
  \q1 G(t_1-t') \q1 G(t_1-t'') +
  \q2 G(t_2-t') \q2 G(t_2-t'') +
  \q3 G(t_3-t') \q3 G(t_3-t'')
  \nonumber \\
  && + \q1 \q2 [G(t_1-t') G(t_2-t'') + G(t_1-t'') G(t_2-t')]
  \nonumber \\
  && + \q1 \q3 [G(t_1-t') G(t_3-t'') + G(t_1-t'') G(t_3-t')]
  \nonumber \\
  && + \q2 \q3 [G(t_2-t') G(t_3-t'') + G(t_2-t'') G(t_3-t')]
  \} \B(t_1-t_2,t_2-t_3)
\end{eqnarray}
or, after Fourier transformation,
\begin{eqnarray}
  \mu_2^{(3)}(\omega',\omega'') &=& (\w'+\w'') G(\w'+\w'') G(\w')
  G(\w'') \sum_{\q1\q2\q3} \U1 \U2 \U3 \q1
  \nonumber
  \\
  && \times
  \{\q1^2 \B(0,0)
  + \q2^2 \B(\w'+\w'',0)
  + \q3^2 \B(\w'+\w'',\w'+\w'')
  \nonumber
  \\
  &&
  + \q1 \q2 [\B(\w',0)+\B(\w'',0)]
  + \q1 \q3 [\B(\w',\w')+\B(\w'',\w'')]
  \nonumber
  \\
  &&
  + \q2 \q3 [\B(\w'+\w'',\w')+\B(\w'+\w'',\w'')]
  \} .
\end{eqnarray}
\end{subequations}
Ratchet effects are related to $\w''=-\w'=\w$, for which Eq.
(\ref{mu.2.3}) follows after usage of momentum conservation.

\section{Details for $T \to \infty$}
\label{app.hT}

Although straightforward, the calculation for the high-temperature
limit requires some care.  For this calculation, it is convenient to
rewrite Eq. (\ref{mu.2.3}) as
\begin{eqnarray}
  \mu_2^{(3)}(-\omega,\omega)
  &=& - \frac i{\eta} \frac 1{\eta^2 \w^2 + m^2 \w^4}
  \sum_{\q1\q2\q3}' \U1 \U2 \U3 \q1
  \iint_0^\infty dt_{12} dt_{23} \F(\w,t_{12},t_{23})
  \exp[-\E(t_{12},t_{23})]
\end{eqnarray}
with $t_{jk} \equiv t_j-t_k$, $t_{13} =t_{12} + t_{23}$,
\begin{subequations}
\begin{eqnarray}
  \F(\w,t_{12},t_{23}) &\equiv& 2\{\q1\q2 [1-\cos(\w t_{12})]
  + \q1 \q3  [1-\cos( \w t_{13})]
  + \q2 \q3 [1-\cos(\w t_{23})]\}
  \nonumber \\ &&
  \times \frac 2\hbar \sin[\frac\hbar2 \q1 G(t_{12}) \q2]
  \frac 2\hbar \sin[\frac\hbar2 (\q1 G(t_{13}) \q3 + \q2 G(t_{23})\q3 )],
\end{eqnarray}
and
\begin{eqnarray}
  \E(t_{12},t_{23}) &\equiv& - [\q1 W(t_{12}) \q2
  + \q2 W(t_{23}) \q3 + \q1 W(t_{13}) \q3]
  = \frac 12 \langle [\q1 x(t_1) + \q2 x(t_2) + \q3  x(t_3)]^2
  \rangle_0 \geq 0 .
\label{def.E}
\end{eqnarray}
\end{subequations}
\end{widetext}
With increasing $T$, the rectified current shrinks since $\E$
increases proportional to temperature.  The dominant contributions
come from small $t_{12}$ and small $t_{23}$. We proceed with an
expansion of $\E$ and $\F$ to extract the leading orders for large
$T$.

In the high-temperature limit, one can neglect the quantum
contribution to $W$ and expands for small times (using $\q2=-\q1-\q3$
because of momentum conservation)
\begin{eqnarray}
  \E&=& \frac {Tm}{\eta^2} \bigg\{
  \frac {(\q1^2+\q3^2)^2}{2\q3^2} \htm^2
  + \frac 13 \q1^2 \frac{\q1+\q3}{\q3} \htp^3
  \nonumber \\ &&
  + O(\htm^3,\htm^2 \htp,  \htm \htp^2,\htp^4) \bigg\}.
\label{E.exp}
\end{eqnarray}
We introduced dimensionless times $\htp$ and $\htm$ via
\begin{subequations}
\begin{eqnarray}
  \gamma t_{12} &\equiv & \htp - \frac{\q1}{\q3} \htm,
  \\
  \gamma t_{23} &\equiv& \htm + \frac{\q1}{\q3} \htp.
\end{eqnarray}
\end{subequations}

To extract the asymptotics for $T \to \infty$, one has to distinguish
the contributions for $\q1/\q3>0$ and $\q1/\q3<0$ (remember that one
needs to consider only $\q1 \neq 0 \neq \q3$).  Because of causality,
the time integrals cover only the quadrant with $t_{12}>0$ and
$t_{23}>0$ in the $(t_{12},t_{23})$ plane.  This quadrant corresponds
to ranges
\begin{subequations}
\label{quadrant}
\begin{eqnarray}
  \htp>0 \textrm{ and }
  -\frac{\q1}{\q3} \htp < \htm < \frac{\q3}{\q1} \htp
  &\textrm{for}&
  \frac{\q1}{\q3}>0,
  \\
  \htm>0 \textrm{ and }
  \frac{\q1}{\q3} \htm < \htp < -\frac{\q3}{\q1} \htm
  &\textrm{for}&
  \frac{\q1}{\q3}<0.
\end{eqnarray}
\end{subequations}
The integrals are transformed via
\begin{eqnarray}
  dt_{12} dt_{23} = \frac{\q1^2 + \q3^2}{\gamma^2 \q3^2} d\htp d\htm .
\end{eqnarray}

For $\q1/\q3<0$, it is sufficient to retain the quadratic term in Eq.
(\ref{E.exp}) since it implies that $\htm \sim T^{-1/2}$.  Then also
$\htp \sim T^{-1/2}$ according to Eq.  (\ref{quadrant}), i.e., $t_{12}
\sim t_{23} \sim T^{-1/2}$.  Since $\F$ is quartic in small times,
\begin{eqnarray}
  \F &=& \frac{\w^2}{\eta^2 \gamma^2}
  (\q1^2+\q3^2)^3 \frac {\q1(\q1+\q3)}{\q3^2}
  \htm^3 ( \frac{\q1}{\q3} \htm-\htp)
  \nonumber \\ &&
  + O(T^{-5/2}),
\end{eqnarray}
the resulting contributions to $\mu_2^{(3)}$ will be of order
$T^{-3}$.  These terms can be neglected in comparison to terms of
order $T^{-17/6}$ which come from $\q1/\q3>0$.

For $\q1/\q3>0$ it is not sufficient to retain the quadratic term in
Eq.  (\ref{E.exp}) since the integral over $\htp$ would diverge.
Thus, one has to include cubic orders in $\E$ which imply that $\htp
\sim T^{-1/3}$. Consequently, the higher order terms not explicitly
written in Eq.  (\ref{E.exp}) can be neglected.  Since $\F \sim \htm^4
\sim T^{-2}$ we now expect $\mu_2^{(3)} \sim T^{-17/6}$.  The proper
expansion of $\F$ up to order $T^{-2}$ now yields
\begin{eqnarray}
  \F &=& \frac{\w^2}{\eta^2 \gamma^2} (\q1+\q3)(\q1^2+\q3^2)^2 \Big\{
  (\q1^2+\q3^2) \frac {\q1}{\q3^2}
  \htm^3 ( \frac{\q1}{\q3} \htm-\htp)
  \nonumber \\ &-&
  \frac {\q1^2}{2 \q3^2} (\q1+\q3)  \htm^2 \htp^3
  \Big\}  + O(T^{-13/6}).
\end{eqnarray}
Thereby it is sufficient to retain even orders in $\htm$ because the
integral over $\htm$ can be extended to all real values [ignoring
condition (\ref{quadrant})] since the quadratic term in (\ref{E.exp})
provides a cutoff that dominates over the condition (\ref{quadrant})
(the errors decay exponentially in $T$).  Therefore, the leading order
$\F \sim \htm^3 \htp \sim T^{-11/6}$ will {\em not} result in a
contribution to $\mu_2^{(3)}$ of order $T^{-8/3}$ since it is odd in
$\htm$.  Performing the time integrals for the remaining terms,
\begin{eqnarray}
  \mu_2^{(3)}(-\omega,\omega)
  &=& - \frac i{\eta^5 \gamma^2} \frac 1{\gamma^2 + \w^2}
  \sum_{\q1,\q3;\q1/\q3>0} \U1 U_{-\q1-\q3} \U3
  \nonumber \\ & \times&
  \frac{\q1^3}{\q3^4}(\q1+\q3)(\q1^2+\q3^2)^3
  \int_0^\infty d\htp \int_{-\infty}^\infty d\htm
  \nonumber \\ & \times&
  \left( \frac{\q1^2+\q3^2}{\q3} \htm^4 - \frac{\q1+\q3}2 \htm^2 \htp^3
        \right) e^{-\E}
\end{eqnarray}
yields Eq. (\ref{mu.hT}).


\begin{thebibliography}{35}
\bibitem[{not({\natexlab{b}})}]{note:y}
\bibinfo{note}{We found it convenient to absorb a factor $1/\hbar$ in the
  definition of $y$ in comparison to Refs. \onlinecite{Fisher+85} and
  \onlinecite{Kleinert}.}
\bibitem[{not({\natexlab{c}})}]{note:FT}
\bibinfo{note}{We employ the usual convention for Fourier transformation,
  $K(\omega)= \int dt \ e^{i\omega t} K(t)$.}
\end{thebibliography}

\end{document}